\documentclass[prb, twocolumn, showpacs, preprintnumbers, amsmath, amssymb]{revtex4}

\usepackage{graphicx}
\usepackage{dcolumn}
\usepackage{bm}

\begin{document}

\preprint{APS/123-QED}

\title{Single electron charging of impurity sites visualized by scanning gate experiments on a quantum point contact}

\author{A. Pioda, S. Ki\v{c}in, D. Brunner, T. Ihn, M. Sigrist, and K. Ensslin}
\affiliation{%
Solid state physics laboratory, ETH Z\"urich, 8093 Z\"urich, Switzerland
}%
\author{M. Reinwald and W. Wegscheider}%
\affiliation{%
Institut f\"ur experimentelle und angewandte Physik, Universit\"at Regensburg, Germany
}%


\date{\today}

\begin{abstract}
A quantum point contact (QPC) patterned on a two-dimensional electron gas is investigated with a scanning gate setup operated at a temperature of 300 mK. The conductance of the point contact is recorded while the local potential is modified by scanning the tip. Single electron charging of impurities induced by the local potential is observed as a stepwise conductance change of the constriction. By selectively changing the state of some of these impurities, it is possible to observe changes in transmission resonances of the QPC. The location of such impurities is determined, and their density is estimated to be below 50 per $\mu$m$^2$, corresponding to less than 1\% of the doping concentration.
\end{abstract}

\pacs{Valid PACS appear here}
\maketitle

\section{\label{introduction}Introduction}

Quantum point contacts (QPCs) are fundamental building blocks of semiconductor nanostructures. Their characteristic property is the quantization of the conductance at low temperatures, observed in very clean samples.\cite{91vanwees, 88wharam} The sensitivity of a QPC to single elementary charges placed in its proximity has been used for charge detection in quantum dots. \cite{96field, 02sprinzak, 03elzerman, 04dicarlo, 05schleser, 05rogge, 06gustavsson} Scanning probe techniques have been employed for the local investigation of QPCs. Among the experimental work are measurements investigating the general behavior of the conductance of QPCs, \cite{02leroy, 00crook, 00crook3} charging of quantum dots close to a QPC, \cite{02crook} and experiments where the probability density inside a QPC \cite{00crook2} and disordered wires \cite{05aoki1, 05aoki2} was mapped. In addition, coherent branched flow of electrons injected from a QPC into a two-dimensional electron gas (2DEG) was observed, \cite{01topinka, 05leroy} and explained in theory. \cite{02he, 03cresti} 

In this paper we apply scanning probe techniques to a QPC patterned with the local anodic oxidation technique on the 2DEG in a GaAs heterostructure. We use the conducting tip of a scanning force microscope as a local gate and measure the conductance of the QPC as a function of tip position. The tip-induced potential locally modifies the microscopic potential landscape close to the constriction. As a consequence, localized impurity sites can be discretely charged, leading to a change in the conductance. We find that the QPC is able to detect such charging events related to localized charges placed at distances up to 1 $\mu$m away from the constriction. The observation of different characteristic functions of the tip-induced potential detected with the QPC, depending on the tip position, confirms that the observed effects are not related to a direct interaction between the scanning tip and the detector, but are mediated by charging at a different location. We also show that there is a direct relation between impurity charging and resonances in the transmission of the QPC. Our data analysis shows that depending on tip position, the influence of certain impurities on the conductance can be changed by the presence of the tip, and transmission resonances can be reduced. Transconductance measurements performed at low magnetic field allow to determine the exact coordinates of impurity sites, and to distinguish between charging and discharging of such sites. A similar behavior in space for different sites suggests that electrostatic coupling between different sites may be of importance. The determination of the number of impurities per area is of importance for transport measurements on semiconductor nanostructures, since charging events may negatively affect the quality of the measured data. We observe a very low density of impurity sites, below 50 per $\mu$m$^2$, indicating that the sample material is of very high quality.

\section{Sample and measurement setup}

\begin{figure*}[t]
\includegraphics[width=13cm]{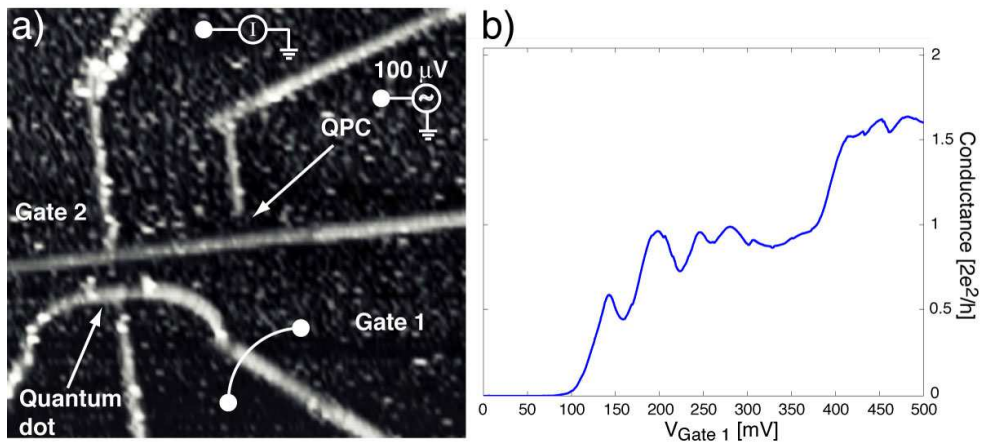}
\caption{\label{fig1}(a)Topography image of the sample taken at a temperature of 300 mK. The QPC used for the measurements is visible in the center of the image. (b) Conductance through the QPC as a function of the voltage applied on the gate region labeled with Gate 1.}
\end{figure*}
The sample has been fabricated on an AlGaAs-GaAs heterostructure containing a two-dimensional electron gas (2DEG) 34 nm below the surface, with density $5\times 10^{11}$ cm$^{-2}$ and mobility 450'000 cm$^2$/Vs at 4.2 K. The nanostructure consists of a quantum dot (not discussed here, see Ref. \onlinecite{04pioda}) and three QPCs. It has been defined by room temperature local anodic oxidation with a scanning force microscope (SFM), \cite{97held, 02fuhrer} which allows to write oxide lines on the surface of a semiconductor, below which the 2DEG is depleted. Figure~\ref{fig1}(a) is a topography image of the sample taken at the base temperature of the measurement setup (300 mK), showing the QPC and the adjacent quantum dot. The conductance of the QPC is controlled by applying a bias voltage to the areas which enclose the quantum dot, labeled ``Gate 1''. The region marked ``Gate 2'' is biased at the same voltage as the QPC region. 

Figure~\ref{fig1} (b) is a conductance trace of the QPC taken with the tip placed at a distant location from the constriction. Several oscillations indicating backscattering in or near the constriction can be observed on and at the onset of the first conductance plateau.

The measurements were performed with a SFM in a $^3$He cryostat with a base temperature of 300 mK. Magnetic fields of up to 9 T can be applied normal to the plane of the 2DEG. The scanning sensor consists of an electrochemically sharpened PtIr tip (with a typical diameter of 30-60 nm for this experiment), mounted on a piezoelectric quartz tuning fork. Details of the low-temperature SFM, as well as of the scanning sensors can be found in Refs. \onlinecite{99rychen} and \onlinecite{04ihn}. The presence of the tip induces a local potential in the 2DEG, whose strength and sign depends on the applied bias voltage. No electron flow between the tip and the 2DEG is possible, since the vacuum gap between the tip and the surface, and the insulating barrier of 34 nm between 2DEG and surface form an insulating barrier. Moving the tip in the vicinity of the QPC leads to changes in its conductance due to the capacitive coupling between the tip and the 2DEG. The scanning gate images presented in this paper are obtained by scanning the tip at a fixed tip-sample separation of about 60 nm and at a fixed tip bias voltage over the area around the QPC. The conductance has been measured either by applying an AC bias voltage of 100 $\mu$V across the QPC and measuring the current with lock-in technique, or by applying an AC bias (in addition to a DC bias) on the scanning tip, and measuring the induced current through the QPC biased with a DC voltage (transconductance). 

From other measurements performed during the same cool-down of the sample (Ref. \onlinecite{05kicin}) it is clear that a double tip is present, with different contact potential differences. SEM images of the tip taken after the experiment confirmed the presence of a particle of different material close to the apex of the tip. The contact potential difference arises from the work function difference between PtIr and GaAs and from the Fermi level pinning at the GaAs surface. The contact potential difference was determined during previous measurements,\cite{04pioda} using the quantum dot as a detector. In that case, the tip is scanned along a line close to the dot for different tip bias voltages. Since the dot is tuned in the Coulomb blockade regime, the conductance resonances observed in the linescan correspond to lines of constant energy. If the position of these lines does not change with tip position, the selected tip bias voltage exactly compensates the contact potential difference. The contact potential difference of one tip is 550 mV, while the other tip has a contact potential difference close to 0 mV. By performing the same linescan measurement on the QPC, the contact potential differences are found to be $V_{CPD}^{tip_{1}}$ = 200 mV and $V_{CPD}^{tip_{2}}$ = -1700 V. They were obtained by determining the tip voltage necessary to keep the QPC conductance constant, when the tip was moved from a distant location to a position on top of the constriction. \cite{06pioda} The measurement is displayed in Fig.~2. The contact potential differences are compensated at a voltage scale, which is outside the area of the plot, and have been obtained by fitting a curve with the same shape as the black parabolae in Fig.~2 at a conductance level corresponding to the background conductance. The measurement was not extended to larger voltages, since care has to be taken not to irreversibly change the electronic properties of the sample.
\begin{figure}[t]
\includegraphics[width=8cm]{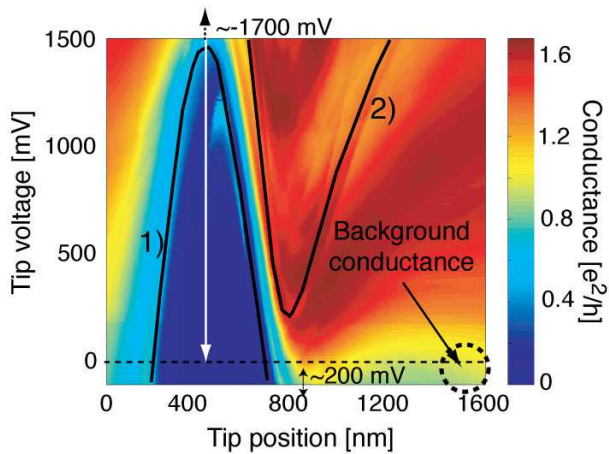}
\caption{\label{fig2} (Color online) Determination of the contact potential difference. The image shows the conductance through the QPC for different positions of the tip and tip bias voltages. The tip was scanned along the dotted line in Fig.~1(a). The contact potential difference is assumed to be zero, when the conductance of the QPC for a certain tip voltage and position corresponds to the conductance in the absence of the tip.}
\end{figure}

\section{Results and discussion}

We identify two different effects of the scanning tip on the conductance of the QPC. The straightforward way to understand the influence of the tip on the QPC is to consider it as an additional gate, which shifts the energies of the quantized modes. This leads to large conductance changes (as already shown in Fig.~\ref{fig2}), and can be modeled in terms of elementary electrostatics. We call this the \textit{gating effect} of the tip. The second effect leads to much more subtle features, which can be barely recognized without further data analysis, and form sets of concentric conductance steps (see Fig.~\ref{fig4}) around different centers. These features can be found up to 1 $\mu$m away from the QPC constriction. We start our discussion of the results with the gating effect.

\subsection{The gating effect}

\begin{figure}[t]
\includegraphics[width=7cm]{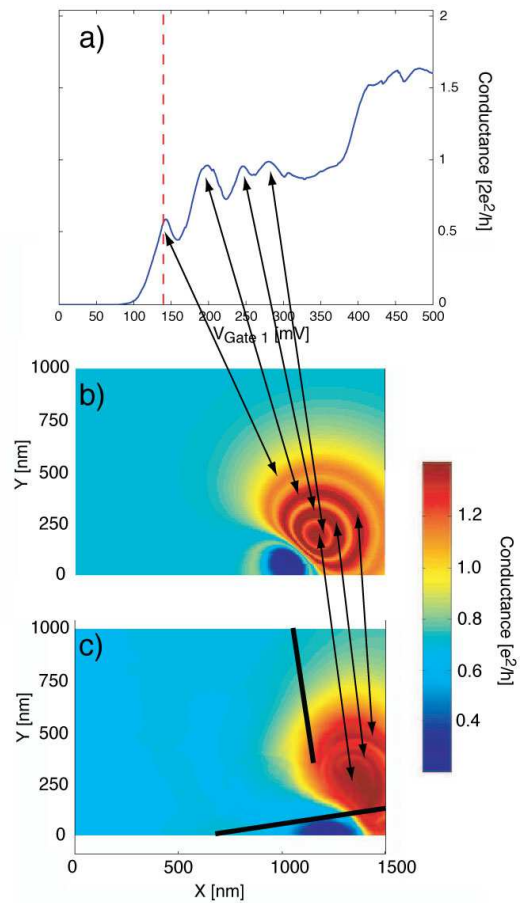}
\caption{\label{fig3}(Color online) (a) Conductance trace of the QPC with the tip placed over the constriction. The voltage applied on gate 1 for the measurement in (c) is marked by the dashed line. (b) Simulation of the gating effect obtained by using the conductance trace and a model for the tip-induced potential (explained in the text) with the same parameters as used for the experiment. (c) Measured scanning gate image showing very similar features as the simulated image. The transmission resonances match almost exactly the experimentally observed ones. The position of the QPC is drawn in the image.}
\end{figure}

A typical scanning gate image is displayed in Fig.~\ref{fig3}(c). The position of the oxide lines defining the QPC constriction obtained from a topography scan has been drawn into the figure for better orientation (thick black lines). The voltage applied on gate 1 is 140 mV, and corresponds to a position in the conductance trace [Fig.~\ref{fig3}(a), dashed line] slightly below the first transmission resonance, while the tip voltage is 1100 mV. The influence of the two different tips in the scanning gate image [Fig.~\ref{fig3} (c)] is evident. 
In the lower right part of the figure, the conductance through the QPC is reduced almost to zero, when the tip with an effective voltage of -600 mV is scanned over the constriction. The remaining right part of the figure shows a strong increase in conductance, with some oscillations, due to the presence of the tip with a (positive) effective voltage of 1300 mV. The oscillations can be matched quite exactly with the transmission resonances observed on the conductance trace in (a). Thus, the tip, acting as an external gate shifts the position of the conductance trace of the QPC to lower or higher gate voltages. It is possible to verify this effect with a model of the tip-induced potential. The exact shape has been reconstructed from the linescan displayed in Fig.~\ref{fig2} and a second linescan crossing it at an angle of 90$^\circ$ on top of the QPC, and can be fitted fairly well with two lorentzian curves, \cite{96eriksson, 06pioda} given by
\begin{equation}
\label{eq1}
\alpha_{i}(r)=\frac{w_{i}^2}{w_{i}^2 + (r-r_{i})^2}
\end{equation}
which correspond to the effective lever arm of the tip $i$ on the QPC for a given position $r_{i}$. The conductance of the QPC for a certain position of the tip can then be written as 
\begin{eqnarray}
\label{eq2}
G_{QPC}[r_{tip}]&=&G\left[ V_{gate 1} + (V_{tip_{1}}-V_{CPD}^{tip_{1}})\alpha_{tip_{1}}(\vec{r}_{tip}) \right.\nonumber \\
&+&\left. (V_{tip_{2}}-V_{CPD}^{tip_{2}})\alpha_{tip_{2}}(\vec{r}_{tip})\right],
\end{eqnarray}
where $V_{tip_{i}}-V_{CPD}^{tip_{i}}$ is the effective voltage on tip $i$. By combining the fit of the tip-induced potential shape with the conductance trace as a function of the voltage on gate 1 [Fig. \ref{fig3}(a)], one obtains Fig. \ref{fig3}(b). The transmission resonances of this image match those of the conductance trace by construction. A comparison with Fig.~\ref{fig3}(c) shows that they also match the resonances observed in the scanning gate measurement. Thus, the changes in QPC conductance described so far are due to the gating effect of the tip on the QPC.

\subsection{Localized charges}

\begin{figure}[t]
\includegraphics[width=8cm]{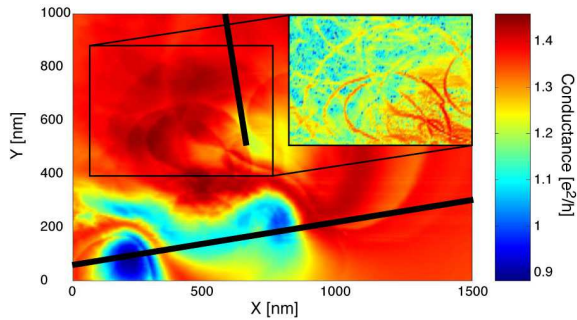}
\caption{\label{fig4}(Color online) Scanning gate image of the area surrounding the constriction (drawn in the image). The inset shows a detail of the region (gradient of the measured data), where charging events are present.}
\end{figure}
The gating effect is very strong, if the QPC shows strong conductance changes for a small change in gate voltage. If the conductance is on a plateau, a shift in voltage induced by the tip will induce only small changes in the conductance of the QPC. An example is shown in Fig.~\ref{fig4}, where the scanning gate image was obtained by scanning the tip with a bias voltage of 1100 mV, and applying a voltage on gate 1 of 250 mV, about in the middle of the first plateau [see Fig.~\ref{fig3} (a)]. The gating effect of the tip is still visible, but also more subtle conductance steps in the region close to the constriction appear in the figure. They become visible, if the measured data are differentiated in scanning direction, as shown in the inset of Fig.~\ref{fig4}. If these conductance changes are related to localized charges and not directly to the conductance of the QPC, then one would expect a different lever arm of the tip on those charges, with respect to the lever arm on the QPC.

\begin{figure}[t]
\includegraphics[width=8cm]{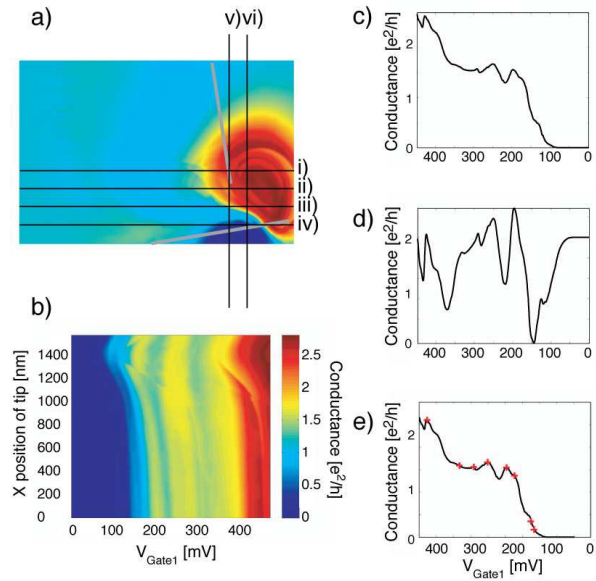}
\caption{\label{fig5}(Color online) Wavelet analysis of the linescans used to determine the lever arm of the tip with respect to local impurity sites. (a) Position of the different linescans and (b) a set of linescans for different gate voltages. A single conductance trace as a function of gate voltage before (c) and after (d) the wavelet analysis. The local maxima of (d) are marked in the trace (e), which is identical to (a) otherwise. The markers are then plotted in conductance grayscale, showing the evolution of local maxima with tip position (see Fig.~\ref{fig6} on the left).}
\end{figure}

In order to separate the effects, linescans have been performed. The measurements are done by scanning the tip along the lines indicated in Fig.~\ref{fig5}(a), and for gate voltages between 0 and 450 mV. The QPC conductance is measured as a function of $V_{gate1}$ for every tip position along the selected line.

\begin{figure*}[t]
\includegraphics[width=16cm]{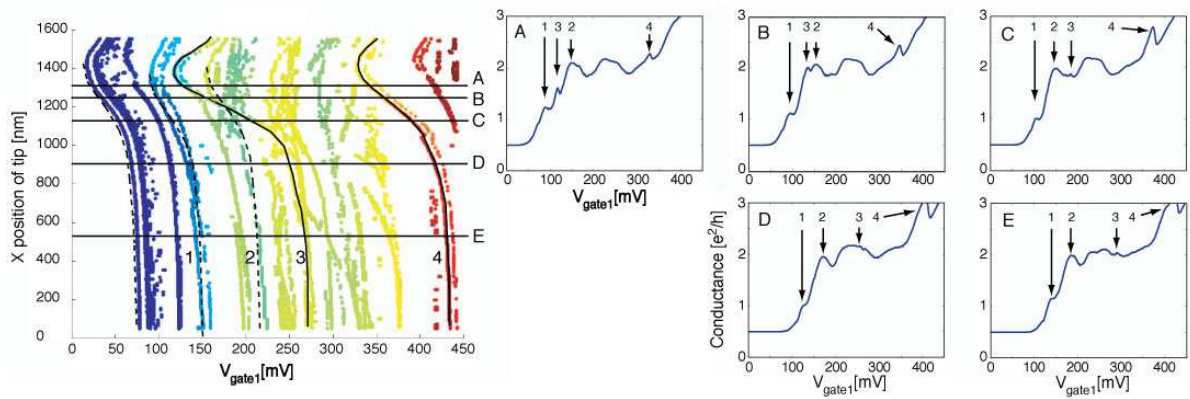}
\caption{\label{fig6}(Color online) Charging events observed as different lever arms of the scanning tip. The left image shows the position of local maxima as a function of tip position and gate voltage (see Fig.~\ref{fig5}(b) for the raw data. The linescan was done on line i) of fig \ref{fig5} (a)). The curves labeled 1-4 mark the position of single local maxima; while curves 1,2, and 4 are parallel to the conductance onset (left dashed line), curve 3 displays a larger change in voltage for different tip positions. Figures A-E on the right are conductance traces obtained from the raw data, and show the position of the local maxima 1-4 for the tip positions denoted in the left figure.}
\end{figure*}
 Since we are interested in finding local maxima of the conductance, which are usually difficult to extract from the measured data directly, filtering algorithms were applied. A simple low- and high pass filtering using Fourier transform did not lead to satisfactory results, since it is hard to define a cut-off frequency, without cutting useful data. The data are therefore filtered using a wavelet analysis, which uses a set of localized base functions instead of a set of plane waves like the Fourier transform. The base functions are generated by scaling and translating a fast decaying function with vanishing average. The base wavelet used here is the Haar wavelet, which is basically a square cosine wave of one period. The filtering procedure works as follows: A complete set of linescans is shown in Fig.~\ref{fig5}(b), and one single line in (c). After applying the wavelet filter, the same line looks as displayed in fig \ref{fig5} (d). The local maxima visible on this trace are marked on trace (e) (red crosses), and subsequently with the correct grayscale in the complete set, as shown in Fig. \ref{fig6} on the left. Fig \ref{fig6} (left) shows that the positions of the local maxima change with tip position following lines parallel to the dotted black line on the left, and lines 1 and 2. These maxima correspond to a single conductance resonance maximum, and are due to the gating effect of the tip, since they are also parallel to the onset where the QPC starts to conduct [light blue region in Fig.~\ref{fig5}(b)]. Some of the lines (e.g. lines 3 and 4), however, follow a different path, indicating that the lever arm of the tip on that specific corresponding resonance is different than for other resonances of the QPC. The lever arm is always larger, up to three times, but never smaller than for the QPC related resonances. The effect is also observable on the conductance traces of the QPC: Most of the resonances shift their position in gate voltage together with the shift in voltage of the overall conductance of the QPC, if the tip position is changed. The resonances related to different lever arms on the other hand tend to migrate from one position to the other, in a way, which is much more extreme than for the other curves. 

\begin{figure}[b]
\includegraphics[width=8cm]{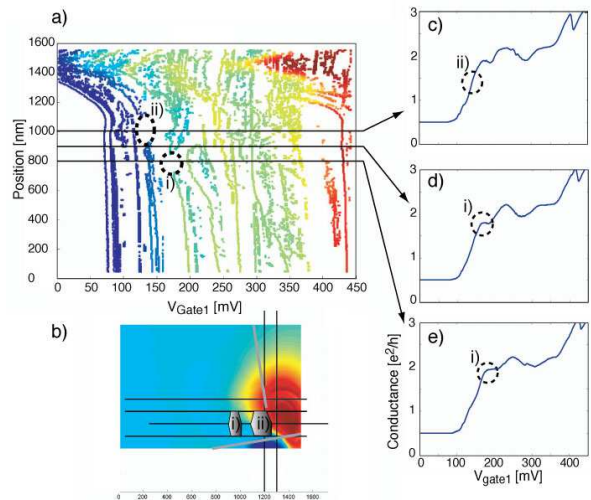}
\caption{\label{fig7}(Color online) Conductance resonances can be made to disappear, if the tip is placed over the position of certain localized charges. a) Position of local maxima after a wavelet analysis (the linescan was done on line iii) of fig \ref{fig5} (a)). The regions marked with i) and ii) lack a maximum. b) Position of the linescans with respect to the QPC. The lack of local maxima can be observed on all linescans at the positions marked i) and i)), thus a scatterer can be expected to be positioned at those locations. c) and e): Conductance traces showing the absence of transmission resonances at the indicated location. In d), the resonance i) is present, but difficult to see.}
\end{figure}

Transmission resonances in QPCs are thought to be caused by a spatially fluctuating potential in the channel or close to the constriction. These fluctuations in the potential can have their origin in a roughness of the constriction, or in individual scatterers placed close to it. Our measurements suggest that localized scatterers can be one of the origins of the observed transmission resonances, and that the scanning tip is able to strongly modify the scattering configuration, leading to position dependent effects of the resonances. 
An additional observation which may support the argument about localized charges, is that single resonances can be made to disappear, if the tip is placed at certain positions. Figure~\ref{fig7}, displays some of these events. The conductance traces c) to d) show a lack of certain resonances, which is also expressed by the sudden disappearance of the corresponding peak line in the plot obtained after applying the wavelet analysis to the measured data [Fig. \ref{fig7} (a)]. By comparing the positions where a local maximum disappears in the different linescans, we observe that they all can be restricted to the two regions marked by i) and ii) in Fig.~\ref{fig7}(b).
These events can therefore be roughly located with respect to the sample, but the small number of linescans does only allow to locate two of them with a relatively low precision.

\subsection{Localized charging in the transconductance}

\begin{figure}[t]
\includegraphics[width=6cm]{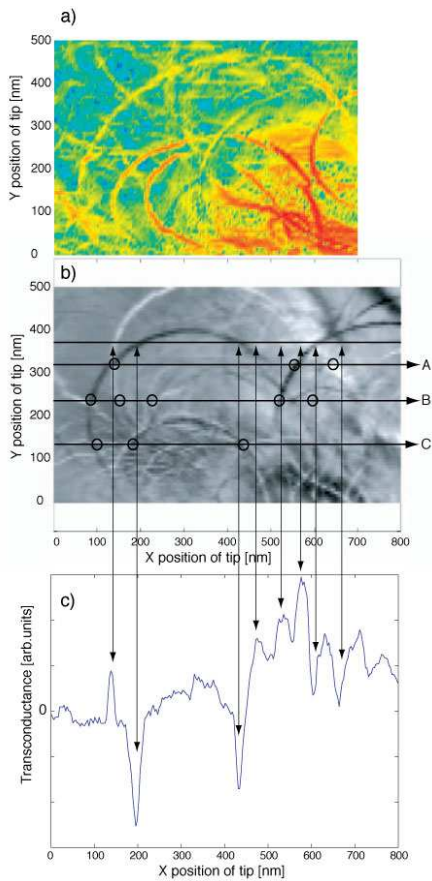}
\caption{\label{fig8}(Color online) Transonductance measurement of the area close to the constriction (b) compared to the DC transport data differentiate mathematically in scanning direction (a) of the inset of Fig.~\ref{fig4}, taken on a different region of the sample. A profile taken through the scanning gate image (c) clearly shows the large current resolution and current peaks corresponding to positive and negative current through the QPC. The lines marked with A-C are discussed in Fig.~\ref{fig9}. The circle mark the crossing points between charging events and the lines A-C.}
\end{figure}

In order to locate more exactly localized charges, we focus on the details shown in the inset of Fig.~\ref{fig4}, which are barely visible in the raw data obtained from scanning gate images. However, a higher resolution can be obtained by measuring the transconductance. In this case an AC voltage of 50 mV at a frequency of 80 Hz is applied to the tip in addition to a DC voltage, and the current through the QPC is detected with lock-in technique at that frequency. Small changes in conductance can be detected with enhanced resolution. By comparing transconductance measurements with the previous conductance data differentiated in scan direction mathematically [Fig. \ref{fig8} (a)], one can see the improvement in resolution. In order to reduce the influence of the transmission resonances, all measurements have been performed in a magnetic field of 2 T. 
A dark ring corresponds to a conductance decrease (or negative current) and thus an electron added to an impurity site, while a light ring corresponds to the opposite situation, where an electron is made to leave the site. Interestingly, no particular ring shows both situations on the same contour, i.e., charging and uncharging events do not appear on the same ring. The profile curve [Fig. \ref{fig8} (c)] clearly shows the different charging events, where a dip corresponds to charging and a peak to uncharging of an impurity site. 

If every ring is assumed to correspond to an impurity site, its position can easily be determined by fitting the contour line with an ellipse, at which center (the center is taken to be the intersection of the main axes) the exact position is expected. By counting the number of centers per area, we obtain a density of about 50 centers per $\mu$m$^2$. This number is very little, if compared to typical doping densities, which are of the order of $10^4$ per $\mu$m$^2$, and thus proves that the sample used is of very high quality. 

\begin{figure}[t]
\includegraphics[width=6cm]{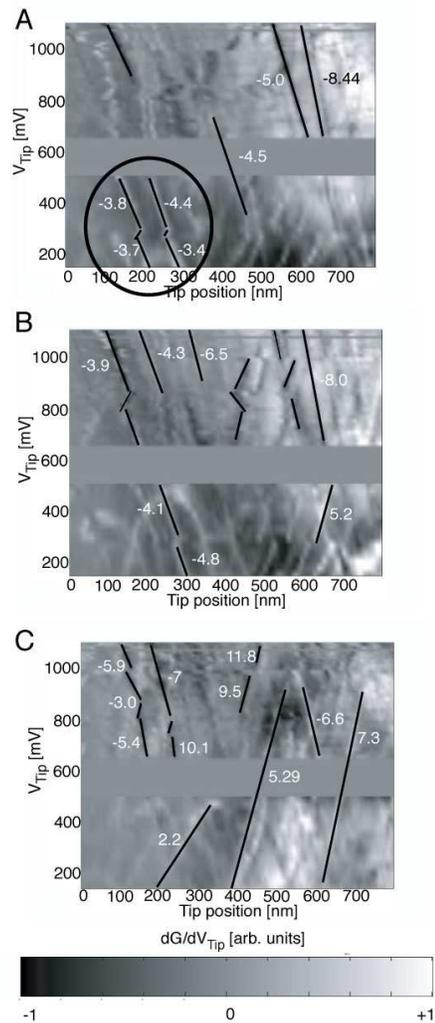}
\caption{\label{fig9}Evolution of the lines corresponding to charging events with tip bias, if the tip is scanned along the lines A-C of Fig.~\ref{fig8}. The numbers give the lever arm of the tip with respect to the lever arm measured from the gating effect. The circle in A highlights a set of charging events which can be either related to sequential charging of the same site, or to a coupling between two different sites.}
\end{figure}
However, already a small density of impurities like those observed here is able to induce changes in the transport current measured through the QPC, in the sense that additional transmission resonances appear, which depend on the charge on such an impurity. 
Considering the low density, they may originate from doping atoms, which are located away from position of the maximum doping density  and whose ground state energy is close to the electrochemical potential of the electrons in the 2DEG. This has to be the case, since the electric field of the tip is effectively screened by the GaAs cap layer, which has a relatively high dielectric constant ($\epsilon=12.9$), and the possible energy changes which can be induced are expected to be below 20 meV from simulations, at the level of the doping plane. 
The presence of such charges in the vicinity of quantum dots could also be an explanation for sudden changes in the conductance due to parametric charge rearrangements. 

The energy dependence of such charging events can be extracted by scanning the tip on a line placed over a region where many impurity sites are observed, and by repeating the scans for different tip voltages. The position of the contour lines is then tracked, as shown in Fig.~\ref{fig9}. The linescans have been performed on the lines shown in Fig.~\ref{fig8}, and marked A--C. By changing the tip bias voltage, the size of the rings changes in a way, which is expected from measurements on quantum dots \cite{04pioda, 05kicin}. We observe linear change in tip bias, and the extracted rate is proportional to the lever arms obtained from the linescans presented above, ranging from 2 to about 12 mV/nm. A rather particular event appears in Fig.~\ref{fig9} A, where a set of two contour lines seems to be coupled (circle in A). This behavior could be due either to two sequential charging events of the same impurity site, or to coupling between different sites, as observed from coupled quantum dots. \cite{96livermore} The first assumption is the most reasonable, since the lever arm of the tip is almost the same for both events, and a small roughness in the confining potential of that specific site could easily lead to a smaller spacing between charging events. Coupling between different sites leading to the anticrossing behavior observed would require two different sites with probably two different lever arms, which are not observed.

\section{Conclusion}

In this paper, we have reported the detection of charging events of impurity sites by a scanning probe at a temperature of 300 mK. The method allows to extract the position of such sites and to determine their density, by using a QPC placed nearby as a sensitive charge detector. By varying the voltage applied on the scanning tip, we were able to determine the energy dependence of the charging events on the position of the tip. We have found that conductance resonances, which are a typical effect of a one-dimensional channel with backscattering, can be influenced by selectively perturbing the charging sites. From this we can conclude that such impurities can in general affect transport measurements in semiconductor nanostructures, and can be related to parametric charge rearrangements often observed in quantum dot measurements. Their low density, on the other hand, and the relatively small changes induced in transport, are a sign of the very high quality of the sample used.



\bibliographystyle{apsrev}


\end{document}